\documentclass[intlimits,twoside,a4paper]{article}

\usepackage[cp1251]{inputenc}

\usepackage[eqsecnum]{cmpj3}

\usepackage{bm}

\issue{2020}{23}{1}{13401}
\doinumber{10.5488/CMP.23.13401}

\title[Optical properties and single-electron states]%
{Optical properties and single-electron states of the
nanosystem that contains three quantum dots}%

\author[I.V.~Bilynskyi, V.B.~Hols'kyi, R.Ya.~Leshko]{I.V.~Bilynskyi, V.B.~Hols'kyi, R.Ya.~Leshko}
\address{Department of Physics, Drohobych Ivan Franko State Pedagogical University, \\
3 Stryiska St., 82100 Drohobych, Ukraine}

\date{Received	October 28, 2019, in final form January 20, 2020}

\begin{document}

\maketitle

\begin{abstract}
The quantum molecule consisting of three quantum dots that forms a triangle
with its centers is studied. The electron wave function in the nanosystem
is written using the linear combination of orbital quantum wells. The
dispersion equation for numerical calculations of the electron stationary states in a
quantum molecule is found. A numerical calculation of the electron energy spectrum in the molecule formed by three quantum dots of a spherical shape is carried out. 
The influence of the nanocrystal size, the distance
between them and the symmetry of the quantum molecule on the electron energy
spectrum is studied. The cases of symmetry of an equilateral and an
isosceles triangle are considered.
\end{abstract}

\section{Introduction}
The energy spectrum of charge carriers in ideal quantum dots (QDs) 
is presented by the set of discrete levels. This is the reason why QDs are often regarded as 
artificial analogs of real atoms. The presence of an atomic-like energy spectrum of charge 
carriers and the possibility of synthesizing such electronic systems under the ``individual order'' 
make QDs heterostructures quite attractive in terms of the creation of semiconductor lasers, 
photodetectors, single-electron and single-photon devices. Using QDs as elementary ``building blocks'', 
one cannot just obtain electronic systems with the least effective dimension, 
but vice versa --- it artificially increases the dimension by creating electron-linked 
single-dimensional chains or two-dimensional layers of QDs \cite{zi}.

If QDs are sufficiently close in the space so that coherent transitions of electrons are possible between 
them through quantum mechanical tunneling, then configurations associated with electrons, the so-called artificial molecules, are formed. 
In such a system, electrons no longer belong to the specific quantum dot,
but form common molecular orbitals which are an analogue of a covalent bond in natural molecules. 
The simplest example of an artificial molecule is the tunneling coupled QDs \cite{zii}. 
It is precisely this coherent two-level system that is being considered at the moment as a quantum bit (qubit) 
of information for a quantum computer. To implement the qubit, it is proposed to use either spin \cite{1} or ``charge'' \cite{2,3,4}
degrees of freedom, but only as the carrier of information (electrons, holes, excitons).

At present, molecules and arrays of bound QDs are the objects of increasing attention of researchers. 
The formation of such objects greatly expands the set of energy parameters and physical properties of nanoobjects. 
For example, the employment of the arrays of coupled QDs can greatly improve the parameters and expand the spectral range of semiconductor QDs lasers \cite{5}.
The formation of the QDs molecules changes the parameters of their exciton states and changes the optical properties. 
Methods of self-organization that provide the production of molecules and arrays of QDs are currently well developed and continue to be improved \cite{6,7}.

Nanosystems consisting of three tunnel-bound QDs or triatomic artificial molecules make it possible to study interesting phenomena associated 
with electrostatics and molecular states of triple quantum molecules (QM). Experimentally, QDs molecules 
are intensively studied in \cite{8,9,10,11,12,13,14}. There can be two~\cite{8}, three \cite{9,10,11,12,13}, four \cite{14}, and more nanocrystals in the QM. 
The geometry of the molecule (three QDs) can be linear \cite{13,14} or triangular \cite{9,10,11,12}. 
It can be symmetric, when the distances between quantum dots are the same and asymmetrical.

The theoretical works have shown that the base model of quantum molecules with three quantum dots is a two-dimensional parabolic quantum well \cite{15,16,17,18}. 
For the purpose of solving the Schr\"{o}dinger equation, the researches used the method of linear combinations of orbital quantum wells \cite{15}, the Gund-Millikan method \cite{16} 
and Heutler-London method \cite{17}. However, these theoretical works do not take into account the finite band offset at the 
boundary and the three-dimensional confinement of particles in QDs.

The present paper is devoted to the study of the influence of the geometry of the three quantum dots molecule on the electron energy 
and on the absorption coefficient of an electromagnetic wave as a function of QD locations in the OQ molecule.

\section{Formulation and solution of the problem}

\begin{figure}[!b]
\centerline{\includegraphics[width=0.5\textwidth]{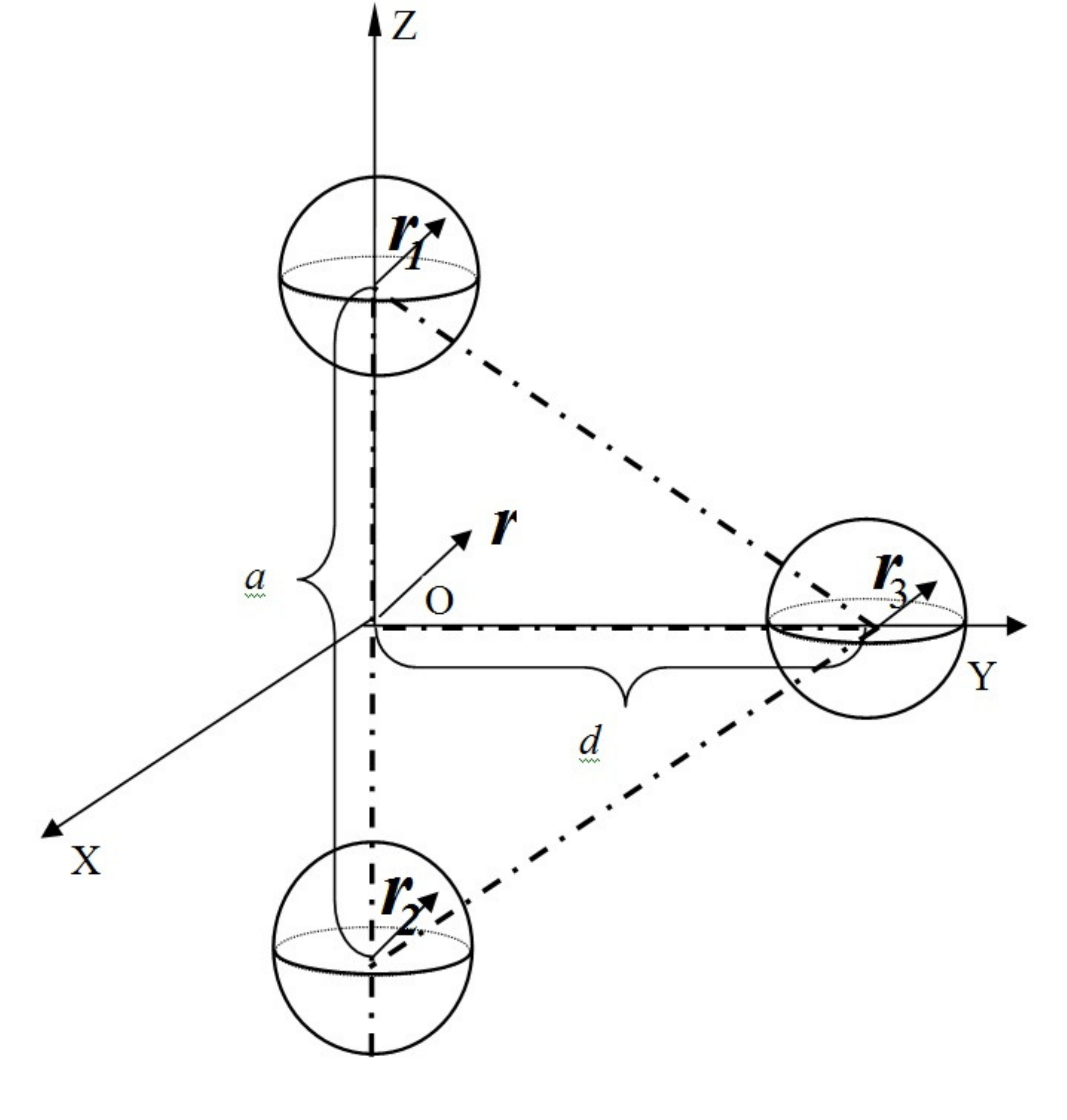}}
\caption{The model of the system.} \label{fig1}
\end{figure}

Let us consider a quantum molecule that consists of three identical spherical QDs. These QDs are coupled due to the tunneling effects. 
We choose the coordinate system so that the centers of two QDs lie on the $OZ$ axis, and the axis origin (point O) lies in the middle between these QDs. 
The third QD is placed on the $OY$ axis at the distance $d$ from the origin (figure~\ref{fig1}).

Let there be one electron in the heterosystem. Then the Hamiltonian of this system in the effective mass approximation is as follows:
\begin{align}
  \label{eq1}
  \hat{H} = - \frac{\hbar^2}{2}\nabla \frac{1}{m({\bf{r}})}\nabla+U({\bf{r}})\,,
\end{align}
\noindent where 
\begin{align}
  \label{eq2}
  U({\bf{r}})=\left\{\begin{array}{l}{0\,, \quad \text { if } {\bf{r}} \text { runs over the matrix, }} \\ {U_{0}\,, \quad \text { if } {\bf{r}} \text { is in any } \text{QD}.}\end{array}\right.
\end{align}

\noindent $U_0 < 0$, $m({\bf{r}})$ is the electron effective mass for the corresponding region:

$$
  m({\bf{r}})=\left\{\begin{array}{ll}{m_{1}\,,} & {\text { if } {\bf{r}} \text { runs over the matrix, }} \\ {m_{2}\,,} & {\text { if } {\bf{r}} \text { is in any } \text{QD}.}\end{array}\right.
$$

To solve the Schr\"{o}dinger equation with Hamiltonian (\ref{eq1}), we use the linear combination of orbital quantum wells approximation. 
The wave function of the electron of the heterosystem can be represented as a linear combination of electron wave functions of individual QD:
\begin{align}
  \label{eq3}
 \Psi({\bf{r}})=\sum_{i=1}^{3} C_{i} \varphi_{i}({\bf{r}}),\,\,\,\,\,\,\, i=1,2,3,
\end{align}
\noindent where $\varphi_i({\bf{r}})$ is the wave function of the ground state of the i-th QD:
\begin{align}
  \label{eq4}
  \varphi_{i}({\bf{r}})=\frac{1}{\sqrt{4 \piup}} \left\{\begin{array}{ll}{A_{i} \frac{\sin k r_{i}}{r_{i}},} & {\text { for } \quad r_{i} \leqslant R,} \\ {B_{i} \frac{\exp \left(-\chi r_{i}\right)}{r_{i}},} & {\text { for } \quad r_{i}>R,}\end{array} \right.
\end{align}
\noindent where
$$
A_i=\frac{1}{\sqrt{\frac{R}{2}-\frac{\sin (2 k R)}{4 k}+\frac{\sin ^{2}(k R)}{2 \chi}}}\,,
$$

$$
B_i=A_i \frac{\sin (k R)}{\re^{-\chi R}},
$$

$$
r_{1}=\left|{\bf{r}}-\frac{{\bf{a}}}{2}\right|, \quad r_{2}=\left|{\bf{r}}+\frac{{\bf{a}}}{2}\right|, \quad r_{3}=|{\bf{r}}-{\bf{d}}|, \ a>2R, \ d>R,
$$

$$
  k=\sqrt{\frac{2 m_{1}}{\hbar^{2}}\left|U_{0}-E^\text{one}\right|}\,, \quad \chi=\sqrt{\frac{2 m_{2}}{\hbar^{2}}\left|E^\text{one}\right|}\,,
$$
\noindent $a$ is the distance between the centers of QD, lying on the $OZ$ axis, $d$ is the distance from another QD to $OZ$ axis, $R$ is the radius of any QD, $E^\text{one}$ is the energy of one isolated QD.

Taking into account (\ref{eq3}) and (\ref{eq4}), the Schr\"{o}dinger equation can be reduced to a system of three algebraic equations:
\begin{align}
  \label{eq5}
  \left\{\begin{array}{l}{C_{1}\left(H_{11}-E\right)+C_{2}\left(H_{12}-E S_{12}\right)+C_{3}\left(H_{13}-E S_{13}\right)=0,} \\ 
  {C_{1}\left(H_{21}-E S_{21}\right)+C_{2}\left(H_{22}-E\right)+C_{3}\left(H_{23}-E S_{33}\right)=0,} \\ 
  {C_{1}\left(H_{31}-E S_{31}\right)+C_{2}\left(H_{32}-E S_{32}\right)+C_{3}\left(H_{33}-E\right)=0,}\end{array}\right.
\end{align}
\noindent where $H_{i j}=\int \varphi_{i}^{*}({\bf{r}}) \hat{H} \varphi_{j}({\bf{r}}) \rd^3{\bf{r}}$, $S_{i j}=\int \varphi_{i}^{*}({\bf{r}}) \varphi_j({\bf{r}}) \rd^3{\bf{r}}$, $i,j=1,2,3$.

The system of homogeneous equations (\ref{eq5}) has a nonzero solution when the determinant of this system is equal to zero:
\begin{align}
  \label{eq6}
  \left|\begin{array}{ccc}{\left(H_{11}-E\right)} & {\left(H_{12}-E S_{12}\right)} & {\left(H_{13}-E S_{13}\right)} \\ 
  {\left(H_{21}-E S_{21}\right)} & {\left(H_{22}-E\right)} & {\left(H_{23}-E S_{23}\right)} \\ 
  {\left(H_{31}-E S_{31}\right)} & {\left(H_{32}-E S_{32}\right)} & {\left(H_{33}-E\right)}\end{array}\right|=0.
\end{align}

Equation (\ref{eq6}) is the dispersion equation for the electron in three QDs molecule.

\section{The energy spectrum of an electron in a molecule of three QDs}

The specific numerical calculations were made for GaAs QDs placed in the AlAs matrix. 
The effective electron mass in these materials is $m_{1}=m_\text{GaAs}=0.063m_0$ and $m_{2}=m_\text{AlAs}=0.15m_0$, where $m_0$ is free electrom mass.
The confinement potential is equal to 560~meV. We consider the QDs of such dimensions that have only one single electronic state. 
This requires 14~\AA$\leqslant R \leqslant$28~\AA. 

It is necessary to determine which parameters of the system affect the electron energy spectrum in the molecule. 
If centers of QDs form a triangle, then their symmetrical placement deserves special attention especially when they 
are vertices of an equilateral triangle. Then the energy spectrum will be influenced by two parameters, i.e., the distance 
between the QDs and their size.

\subsection{Quantum dots are vertices of an equilateral triangle}
\begin{figure}[!b]
\centerline{\includegraphics[width=0.65\textwidth]{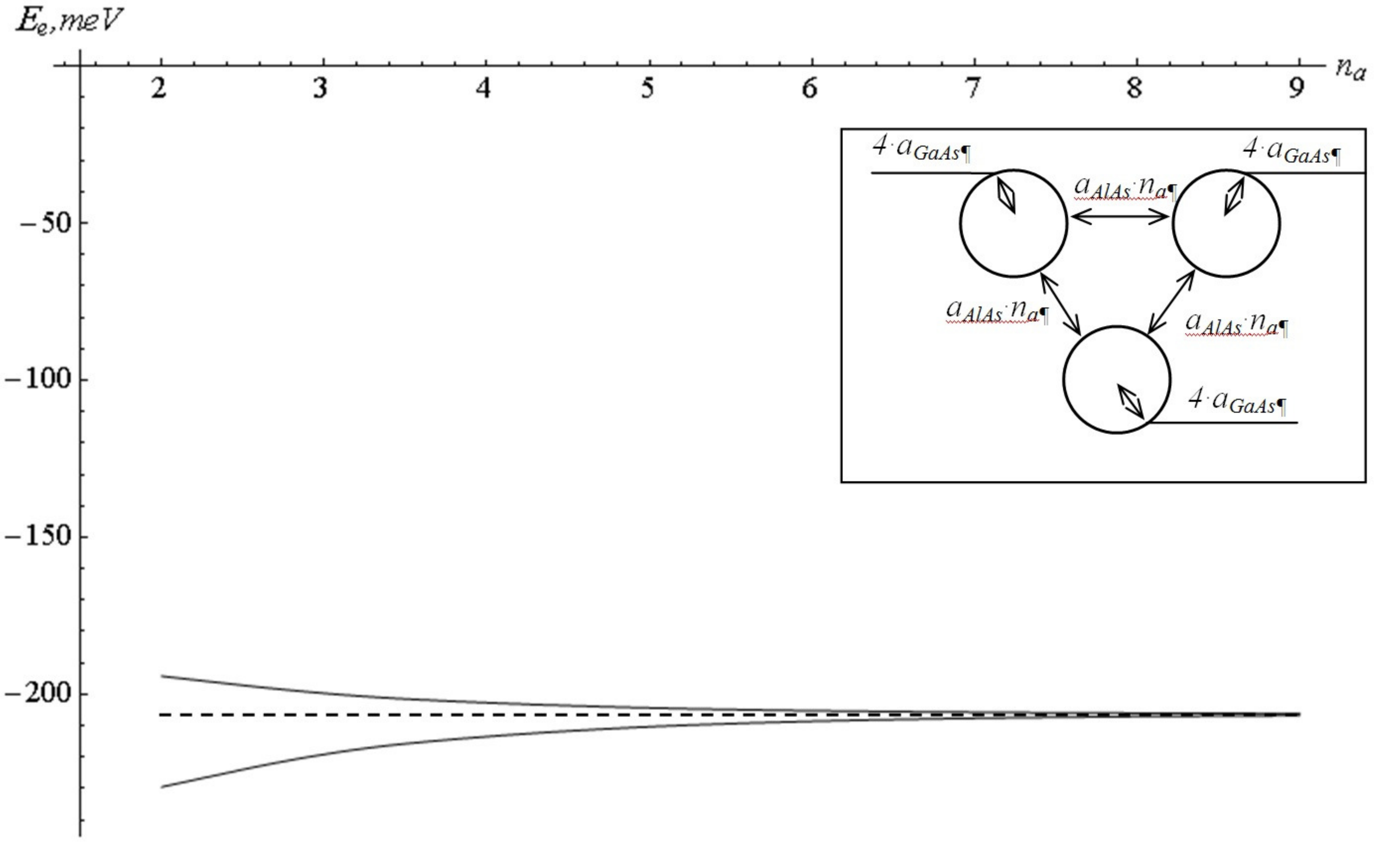}}
\caption{The electron energy in the QM as a function of the distance between nanocrystals with symmetry of 
the equilateral triangle (the distance between QDs in a lattices constants AlAs, and their radii are 
equal to 4 lattices constants GaAs). Dashed line denotes the energy level of an isolated QD.} \label{fig2}
\end{figure}

The dependence of the stationary states of the electron on the distance $a=2R+n_{a}a_\text{AlAs}$, (where $a_\text{AlAs}=5.6611$~\AA, $n_a = 2, 3, \ldots)$ between the centers of the QDs 
of the same radius $R=4a_\text{GaAs}$, ($a_\text{GaAs}=5.6532$~\AA) is shown in figure~\ref{fig2}. It is seen that the energy spectrum of an electron is determined 
by two energy levels obtained as a result of the splitting of the electron level in an isolated nanocrystal. Moreover, the ground state is not degenerate, 
but the first excited state is twice degenerate. If the distances between QDs boundaries is 45.3~\AA~($n_a=8$), then the states merge into one, the value of 
which equals the magnitude of the energy of the electron bound state in an isolated QD of the same size. 

If the distance decreases, then the magnitude of the splitting increases. Thus, for a distance of 11.3~\AA~($n_a=2$), the difference between levels 
is $\Delta E_1=35.3$~meV. In this paper, we also study this heterostructure numerically using the finite element method in COMSOL Multiphysics. 
This method yields three energy levels, two of them being identical, for all considered $n_a$. 
In addition, the energy levels found using the numerical method and the linear combination of orbital quantum well approximation are different by no more than 2\%.

\begin{figure}[!t]
\centerline{\includegraphics[width=0.65\textwidth]{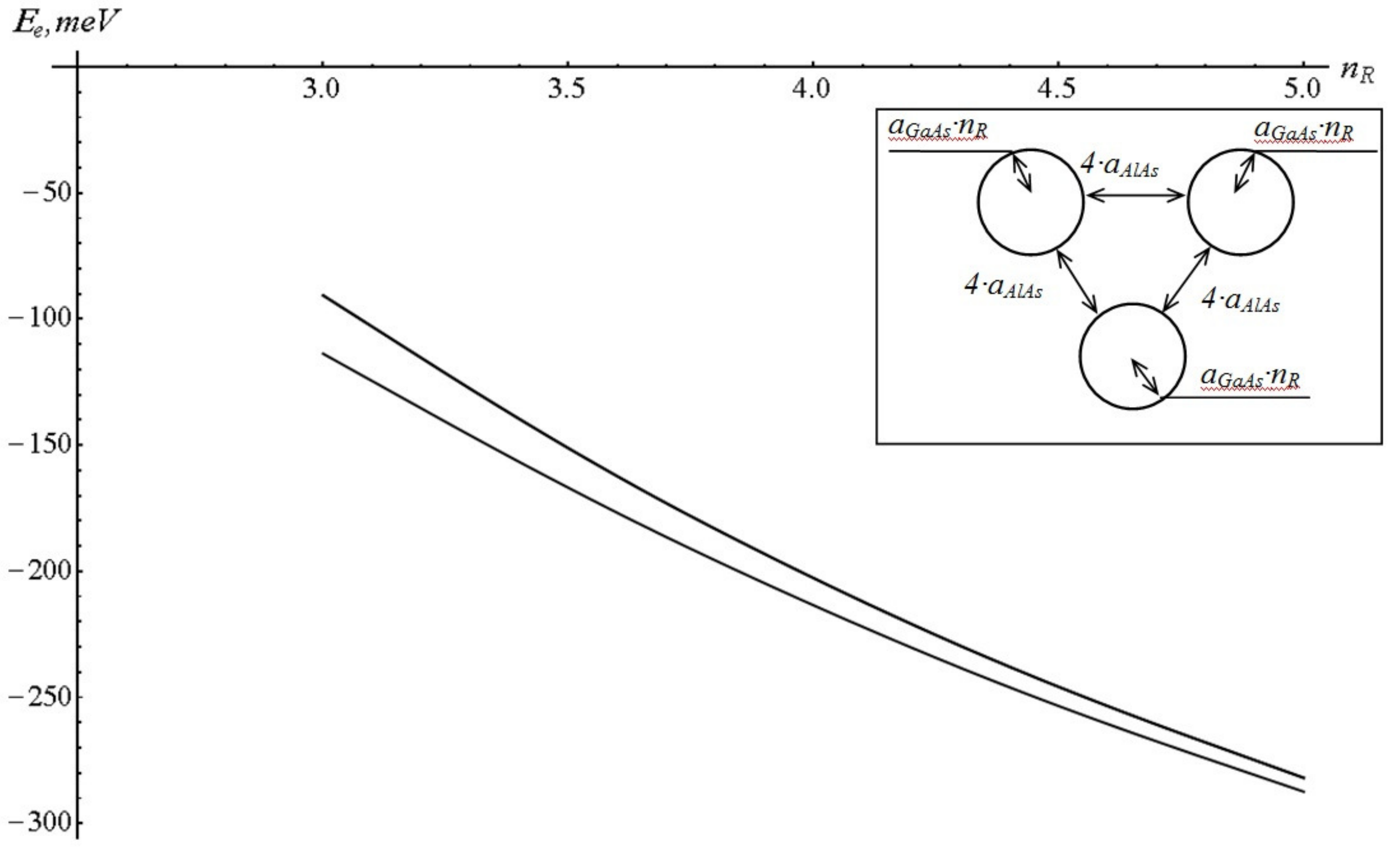}}
\caption{The electron energy of QM as the function of the nanocrystal size in the symmetry of an equilateral triangle 
(the distance between QDs in a lattices constants AlAs, and their radii are equal to 4  lattices constants GaAs).} \label{fig3}
\end{figure}

In figure~\ref{fig3}, we depict the energy of electron stationary states of a molecule of QDs for different radii of QDs ($R=n_R a_\text{GaAs}$) 
if the distance between their boundaries does not change and is equal to 22.64~\AA~(4$a_\text{AlAs}$). As we can see, if the size of nanocrystals increases, 
then the magnitude of the energy splitting decreases. Thus, at $R=16.96$~\AA~($n_R=3$), the splitting value is 23.1~meV. For the value of $R$, which corresponds to 5 lattice constants of GaAs, it is equal to 5.5~meV.

\subsection{Quantum dots are the vertices of an isosceles triangle}
\begin{figure}[!t]
\centerline{\includegraphics[width=0.65\textwidth]{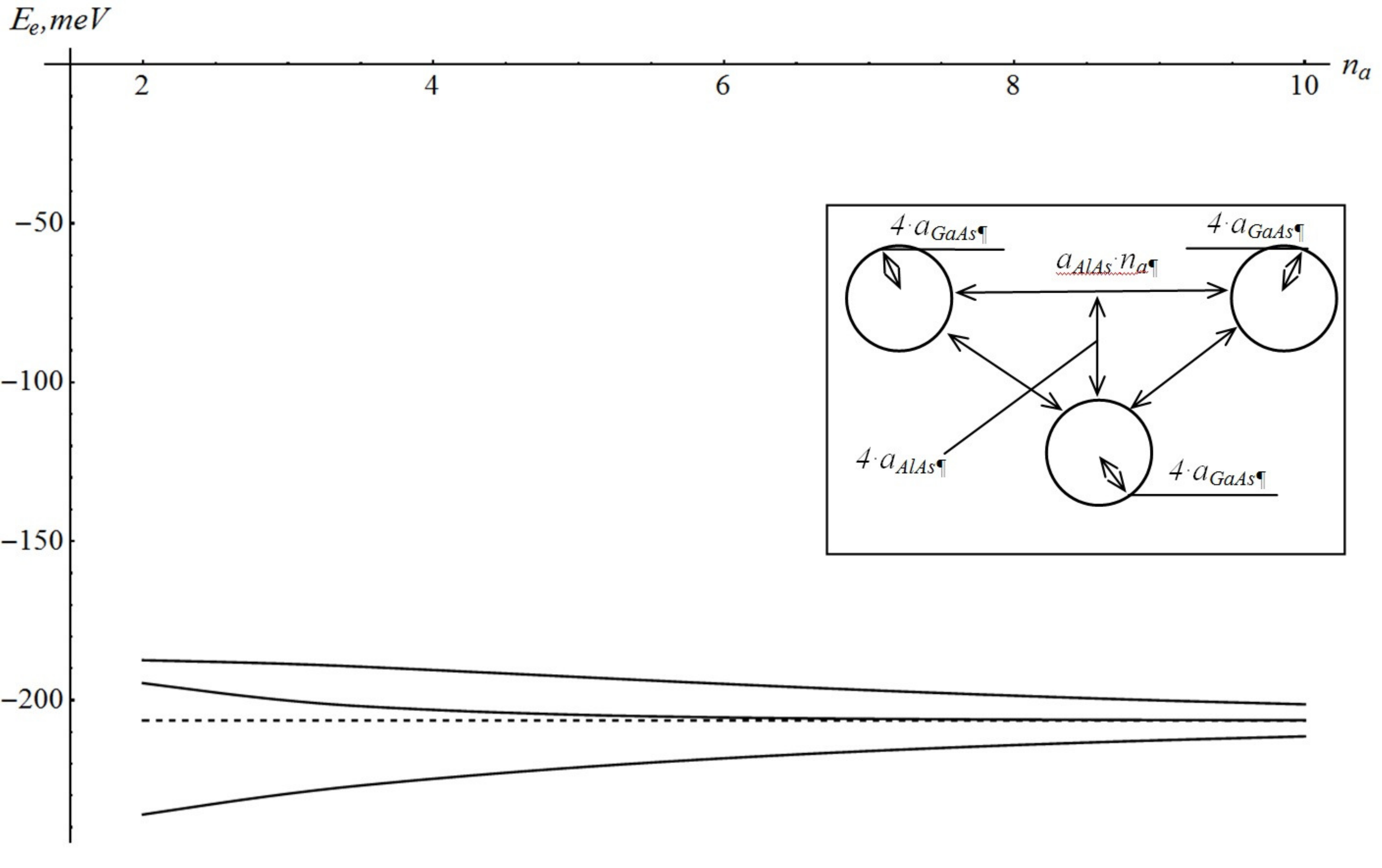}}
\caption{The electron energy in the QM as a function of the distance between two QDs when the third QD 
is at the same distance to the axis of the other two at the symmetry of the isosceles triangle 
(the distance from the third QD to the axis is 4 lattices constants AlAs, and their radii are equal to 4 lattices constants GaAs).
Dashed line denotes the energy level of an isolated QD.} \label{fig4}
\end{figure}

Let us now consider the other geometry of the three quantum dots molecule. 
Let the triangle, which is formed by the centers of QDs, be isosceles, provided that the altitude $d$ (figure~\ref{fig1}) is the same. 
Figure~\ref{fig4} shows the effect of changing the distance on the electron energy, provided $d=4a_\text{AlAs}$, $R=22.6$~\AA. 
In this case, the heterostructure is characterized by three energy levels, which is explained by the decrease in the spatial symmetry of the molecule.
Even at large distances, there is a small split: it is about 5~meV between the nearest states. 
The energy levels of the excited states monotonously decrease with an increasing distance $a$. 
The magnitude of the energy splitting between the highest and lowest levels at the distance of 11.3~\AA \ is equal to $\Delta E_1=48.5$~meV, 
and between the ground and the first excited state is $\Delta E_2=41.3$~meV, and between excited states is $\Delta E_3=\Delta E_1-\Delta E_2=$7.2~meV.

\begin{figure}[!t]
\centerline{\includegraphics[width=0.65\textwidth]{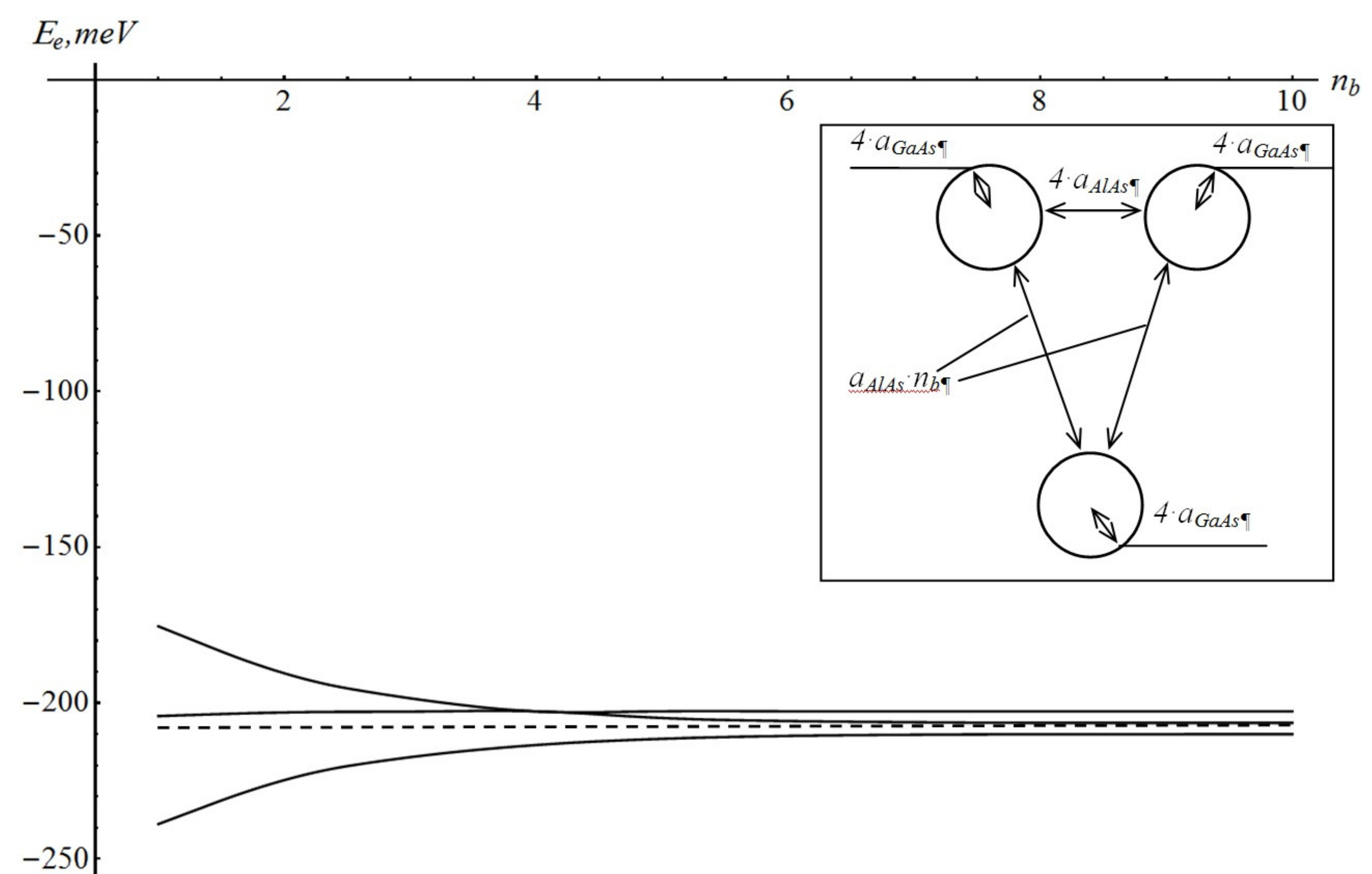}}
\caption{The electron energy the QM as a function of the distance of one QD to the axis on 
which there are two other QD at the symmetry of an isosceles triangle 
(the distance between two QDs is equal to 4 lattices constants AlAs, and the radii of all levels are 4  lattices constants GaAs).
Dashed line denotes the energy level of an isolated QD.} \label{fig5}
\end{figure}

Figure~\ref{fig5} shows the dependence of the electron energy in a molecule formed with three QDs as function of $n_b~(n_b=1,2,\ldots)$ 
at the fixed distance $a$ between the centers of two QDs ($a=67.87$~\AA). The sizes of nanocrystals are the same as in the previous case ($R=22.6$~\AA).
On the abscissa, the distance $n_b$ (in lattice constants $a_\text{AlAs}$) to the bounds of the other two QDs is presented. 
It is seen that the ground energy level behaves the same way as in the previous case. 
For excited levels at $d=58.8$~\AA, there is observed a degeneration, because in this case we have an equilateral triangle from the centers of QDs. 
It should also be noted that the energy of one level is practically unchanged. However, it transforms from the first excited level into the second one (the order of levels is changed).
The magnitude of the same splitting with a distance $d=11.3$~\AA \ between the highest and the lowest is equal to $\Delta E_1=34.2$~meV, 
between the ground and first excited is $\Delta E_2=21.7$~meV, and between the excited states is $\Delta E_3=12.5$~meV. 

\section{The light absorption coefficient of the heterosystem with quantum molecules}

Let us find the form of the wave function of a particle in the heterostructure with three QDs located at the vertices of the triangle. 
To do this, we use the system of equations (\ref{eq5}). The wave function (\ref{eq3}) can be rewritten in the form:
\begin{align}
  \label{eq7}
  \Psi({\bf{r}})=\left[\alpha_{1} \varphi_{1}({\bf{r}})+\alpha_{2} \varphi_{2}({\bf{r}})+\varphi_{3}({\bf{r}})\right] C_{3}\,,
\end{align}
\noindent where

$$
\alpha_{1}=-\frac{\alpha_{2}\left(H_{12}-E S_{12}\right)+\left(H_{13}-E S_{13}\right)}{\left(H_{11}-E\right)},
$$

$$
\alpha_{2}=\frac{\left[\left(H_{13}-E S_{13}\right)\left(H_{21}-E S_{21}\right)-\left(H_{23}-E S_{23}\right)\left(H_{11}-E\right)\right]}{\left[\left(H_{22}-E\right)\left(H_{11}-E\right)-\left(H_{12}-E S_{12}\right)\left(H_{21}-E S_{21}\right)\right]}.
$$

Using the normalization condition, we find the coefficient $C_3$:
\begin{align}
\label{eq8}
C_{3}=\frac{1}{\sqrt{\int\left|\alpha_{1} \varphi_{1}({\bf{r}})+\alpha_{2} \varphi_{2}({\bf{r}})+\varphi_{3}({\bf{r}})\right|^{2} d^3{\bf{r}}}}.
\end{align}

The light absorption coefficient caused by the interlevel transition is determined on the basis of the formula \cite{19,20}:
\begin{align}
  \label{eq9}
  \alpha(\omega)=\omega \sqrt{\frac{\mu_{0}}{\varepsilon_{0} \varepsilon}} \frac{\sigma\left|\rho_{1,2}\right|^{2} \hbar \Gamma_{1,2}}{\left(E_{2}-E_{1}-\hbar \omega\right)^{2}+\left(\hbar \Gamma_{1,2}\right)^{2}}\,,
\end{align}
\noindent where $\omega$ is a wave frequency, $\varepsilon_0$ is an electric constant, $\mu_0$ is a magnetic constant, $\varepsilon$ is a dielectric
permeability of the QD, $\hbar \Gamma_{1,2}$ is a relaxation energy due to the electron-phonon interaction and other scattering factors 
(for calculations in this work it was taken $\hbar \Gamma_{1,2}=5$~meV \cite{3,4}), $\sigma$ is density of quantum molecules in the matrix;  
$\rho_{1,2}$ is the matrix element of dipole momentum between states $|1\rangle$ and $|2\rangle$.

\begin{figure}[!t]
\centerline{\includegraphics[width=0.65\textwidth]{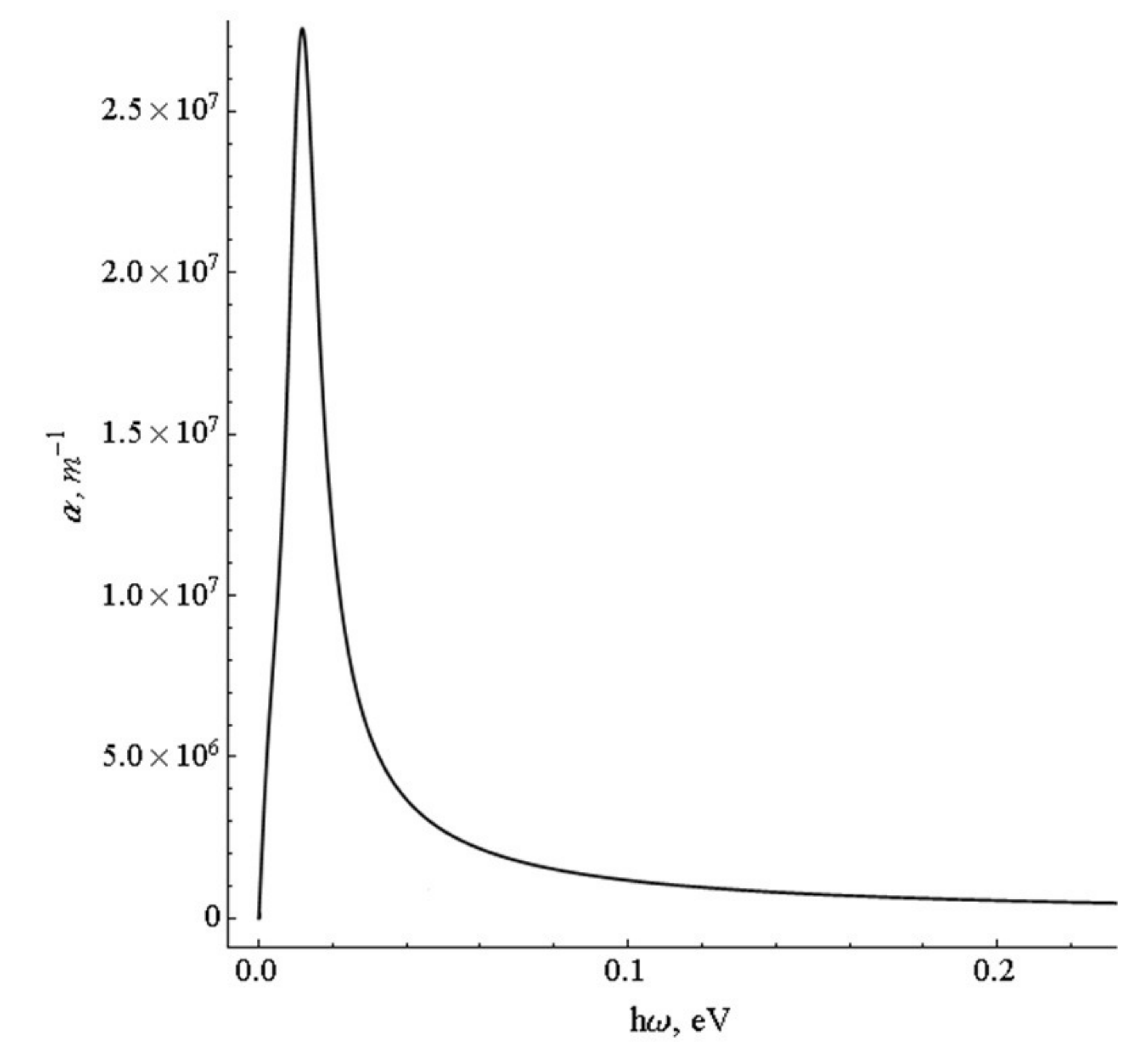}}
\vspace{-2mm}
\caption{Dependence of the electromagnetic wave absorption coefficient on the quantum energy of the light for the 
case of an equilateral triangle (linear-polarized light: the vector polarization is directed 
along the altitude of the triangle).} \label{fig7}
\end{figure}
\begin{figure}[!t]
	\centerline{\includegraphics[width=0.65\textwidth]{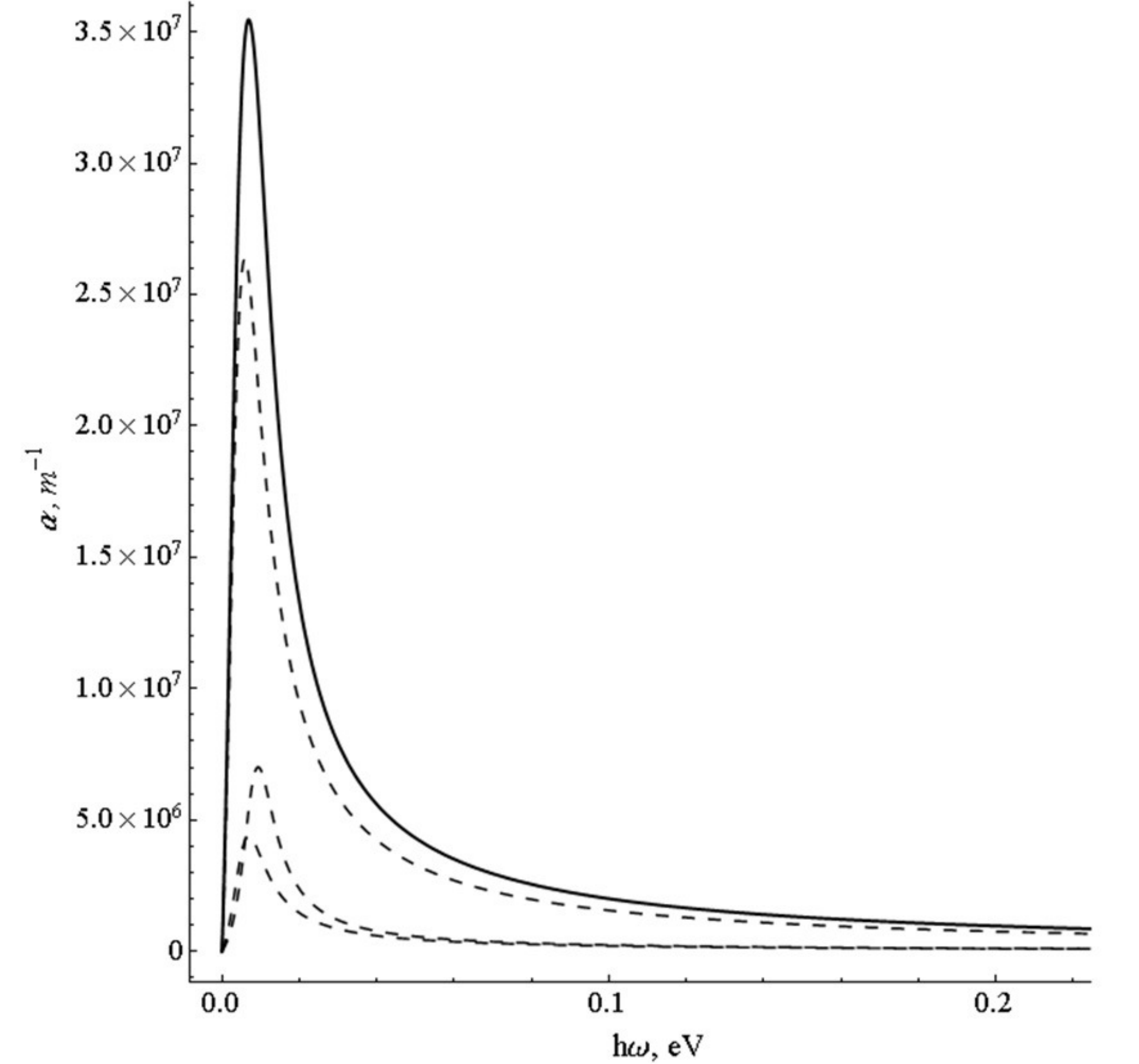}}
	\vspace{-2mm}
	\caption{Dependence of the electromagnetic wave absorption coefficient on the quantum energy of the light for 
		the case of symmetry of an isosceles triangle (linearly polarized light: the vector of polarization is directed along the altitude of the triangle).} \label{fig8}
\end{figure}

In order to determine the absorption coefficient, the linearly and circularly polarized electromagnetic waves were considered. In figure~\ref{fig7} the dependence of the absorption coefficient on the energy of the incident 
quantum for the case of symmetry of an equilateral 
triangle is presented. The system is characterized by one peak that is responsible for the transition from the ground state to the excited one. 
The symmetry of an isosceles triangle is considered in figure~\ref{fig8}. Here, the system is characterized by three states. 
For the polarization vector, which is directed along the altitude of the triangle (figure~\ref{fig8}), three transitions are possible: 
between the ground and excited states and between the excited states. The highest peak is responsible for the transition between the excited states, 
and the smallest is between the ground and the first excited state. If polarization vector of electromagnetic wave is directed along the other axes, the corresponding 
absorption coefficient is very small. For circularly polarized light, the absorption coefficient is very small too.
When the distance between QDs increase, the absorption coefficient (caused by the transitions between the denoted levels) decreases due to decreasing the level splitting.
The presented model can be expanded for the calculation of other forms of QD structures \cite{21}.

\section{Conclusions}
The molecule of three quantum dots placed at the vertices of a triangle has been studied. 
The form of the wave function of the electron in the investigated nanosystem was written using the linear combination of orbital quantum wells. 
A dispersion equation for numerical calculations of stationary states of an electron in a molecule was found. 
The numerical calculation of the energy spectrum of an electron in a molecule, which is formed of three QDs of a spherical shape, was carried out.
The influence of the size of the nanocrystal, the distance between them and the symmetry of the QM on the energy spectrum of the electron was investigated. 
The case of symmetry of an equilateral and an isosceles triangle is considered. 
In order to compare the results, the finite element method was also used to calculate the energy spectrum of an electron in a system of three spherical QDs.
Based on the electron energy and the wave function, the absorption coefficient of the electromagnetic wave in a QM from three spherical QDs was investigated.

\section*{Acknowledgements}

The authors are grateful to our scientific mentor, \fbox{Professor Vasyl~Boichuk} for the idea (expressed by him 2 years ago) of writing this article.

\ukrainianpart

\title{Одноелектронні стани та коефіцієнт поглинання в молекулі, що утворена з трьох квантових точок}
\author{І.В. Білинський, В.Б. Гольський, Р.Я. Лешко}
\address{Кафедра фізики, Дрогобицький державний педагогічний університет імені Івана Франка,  \\ вул. Стрийська, 3, 82100 Дрогобич, Україна}

\makeukrtitle

\begin{abstract}
\tolerance=3000%
Досліджено квантову молекулу з трьох точок, що своїми центрами утворюють трикутник. Методом лінійної комбінації орбіталей квантових ям записано вигляд хвильової функції електрона в досліджуваній наносистемі. Знайдено дисперсійне рівняння для чисельного обчислення стаціонарних станів електрона в квантовій молекулі. Проведено чисельний розрахунок енергетичного спектра електрона в молекулі, що утворена з трьох квантових точок сферичної форми. Досліджено вплив величини нанокристала, відстані між ними та симетрії квантової молекули на енергетичний спектр електрона. Розглянуто випадок симетрії рівностороннього та рівнобедреного трикутника.

\keywords молекула з квантових точок, три квантові точки

\end{abstract}

\end{document}